\newcommand{\BIBFILEYES}[1]{}
\newcommand{\BIBFILENO}[1]{#1}
\newcommand{\WITHPRIOR}[1]{#1}
\newcommand{\WITHOUTPRIOR}[1]{}

\newcommand{\meanrho}{\bar{\rho}}

\newcommand\E{\mathop{\mathbb{E}}\nolimits}

\newcommand{\Qraymond}{\chi}
\newcommand{\chiraymond}{q}
\newcommand{\mraymond}{m}

\newcommand{\mconj}{{\hat \mraymond}}
\newcommand{\Qconj}{{\hat \Qraymond}}
\newcommand{\chiconj}{{\hat \chiraymond}}
\newcommand{\wioversqrtQi}{u_i}

\newcommand{\source}{x}

\newcommand{\cut}[1]{}

\documentclass[conference]{IEEEtran}

\ifCLASSINFOpdf
\else
  \usepackage[dvips]{graphicx}
 \graphicspath{{./}}
 \DeclareGraphicsExtensions{.eps}
\fi

\usepackage[cmex10]{amsmath}

\usepackage{amsfonts}

\begin{document}
\title{Optimal incorporation of sparsity information by weighted $\ell_1$ optimization}
\author{
\IEEEauthorblockN{Toshiyuki Tanaka}
\IEEEauthorblockA{Graduate School of Informatics \\
University of Kyoto\\
Kyoto, Japan\\
Email: tt@i.kyoto-u.ac.jp}
\and
\IEEEauthorblockN{Jack Raymond}
\IEEEauthorblockA{Department of Physics\\
Hong Kong University of Science and Technology\\
Hong Kong, China\\
Email: jackraymond@physics.org}
}

\maketitle

\begin{abstract}
Compressed sensing of sparse sources can be improved by incorporating prior knowledge of the source. In this paper we demonstrate a method for optimal selection of weights in weighted $\ell_1$ norm minimization for a noiseless reconstruction model, and show the improvements in compression that can be achieved.
\end{abstract}

\IEEEpeerreviewmaketitle

\section{Introduction}
Compressed sensing research has the aim of improving the efficiency and reliability of source estimation when the number of measurements taken is smaller than the number of degrees of freedom in the source. Although somewhat counterintuitive, it turns out that, given some special structure in the source, it is possible to reconstruct accurately all source components~\cite{Donoho:CS}.

Owing to the fundamental nature of the problem, it is not surprising that research and methodology span a number of fields such as image processing, topology, multi-user detection, and convex optimization~\cite{Parvaresh:RSS,Xu:BT,Donoho:CF,Fletcher:OO,Candes:DLP}. In this article we also apply methods developed initially within physics, that have recently been applied to compressed sensing~\cite{Kabashima:TR,Rangan:AA,Nishimori:SP}.

In the canonical model, a real-valued vector $\mathbf{y}$ constitutes observations of a real-valued source vector $\mathbf{x}^0$ via a measurement matrix $A$ as
\begin{equation}
\mathbf{y} = A \mathbf{x}^0 \label{eq:yAx}\;.
\end{equation}
The dimension of $\mathbf{y}$ is $M$ and the dimension of $\mathbf{x}^0$ is $N (\geq M)$.
The parameter $\alpha=M/N$ quantifies the number of observations per the degree of freedom of the source, and we call it the compression rate. The problem of compressed sensing is to reconstruct the source $\mathbf{x}^0$ from the observations $\mathbf{y}$ and the measurement matrix $A$.

When the compression rate $\alpha$ is smaller than one, the problem amounts to solving an ill-conditioned set of linear equations and hence its solution is not unique. However, if the source is suitably structured, then such structural information may be utilized to overcome the ill-conditionedness. One example is sparsity, whereby the source has many zero components. For reconstruction of sparse sources it is now well known that provided $\alpha$ is not too small an accurate reconstruction of the source $\mathbf{x}^0$ may be achieved by minimizing the $\ell_1$ norm
\begin{equation}
\left\| \mathbf{x} \right\|_1 = \sum_i |\source_i| \label{eq:L1norm}
\end{equation}
subject to the constraint $\mathbf{y}=A\mathbf{x}$. This minimization problem can be solved efficiently by casting it into a problem of linear programming.
It has recently been shown that $\ell_1$ minimization also allows perfect reconstruction of sparse sources~\cite{Donoho:CF,Kabashima:TR} with high probability for large systems provided that the compression rate $\alpha$ is
larger than a certain threshold value $\alpha_c$. Furthermore a sharp transition between perfect and imperfect reconstruction is observed at $\alpha=\alpha_c$, which we call the compression threshold.

We shall consider weighted norms, in which each component in the sum (\ref{eq:L1norm}) has a different coefficient (weight). The consideration of non-uniform weights has been motivated by Bayesian and topological arguments, algorithmic studies and much empirical evidence.  Numerous results indicate that use of non-uniform weights can reduce the compression threshold, guarantee optimality in a probabilistic sense, or otherwise improve algorithmic performance ~\cite{Candes:ES,Cevher:LCP,Wipf:SE,Wipf:IR}. A topology based method able to place bounds on weighted reconstruction was recently proposed, and demonstrated for the case of a source with two blocks of distinct density of non-zero components~\cite{Xu:BT}.

\section{Problem}

\subsection{Reconstruction with marginal density information}
The problem we address is the following: given a mean density of non-zero components, and some additional prior information on the distribution of non-zero components, how can the weights be chosen so as to minimize the number of measurements required to reliably reconstruct the source using weighted $\ell_1$ norm (w-$\ell_1$) minimization.

We assume a standard model framework in which components of $A$ are independent and identically-distributed random variables; drawn from a symmetric distribution of finite variance and higher-order moments without finite probability mass at the origin, the Gaussian distribution being one intuitive case. The observations $\mathbf{y}$ are also random variables, determined as functions of the measurement matrix and a random source through (\ref{eq:yAx}).

In the standard sparse problem a fraction $0<\meanrho<1$ of all source components are non-zero, this subset being selected uniformly at random. It is instead assumed in this article that a (potentially distinct) marginal probability is known for every component in the source, so that $x_i^0$ is non-zero with probability $0\leq\rho_i\leq 1$ independently of other components. We call the parameter $\rho_i$ the density of component $i$. The prior is therefore
\begin{equation}
P(\mathbf{x}^0)= \prod_{i=1}^N \left[\rho_i S(x_i^0) + (1-\rho_i)\delta(x_i^0)\right] \label{eq:trueprior}\;,
\end{equation}
where $S(x_i^0)$ is the conditional prior of $x_i^0$ given that it is non-zero. The conditional prior $S(x_i^0)$ is assumed unknown to the observer, but it should have finite moments and should not have finite probability weight at $x_i^0=0$.
\WITHOUTPRIOR{We choose a normal distribution in representing results for convenience.} The densities $\{\rho_i\}$ encode prior information about sparsity of the source, are assumed available to the observer, and will be called marginal density information. This definition covers the case of block-structured sources, wherein a source is divided into sets of components according to their densities, including two blocks as a special case.

To each source component is also associated a weight $w_i>0$. The estimate of the source is obtained as a solution to an optimization problem: minimization of the weighted $\ell_1$ norm
\begin{equation}
\min_{\mathbf{\source}}\left( \sum_{i} w_i|\source_i|\right)\qquad\hbox{subject to}\qquad \mathbf{y}=A\mathbf{\source}\;. \label{eq:wL1min}
\end{equation}

We note irrespective of weighting a trivial lower bound on the compression required for perfect reconstruction: $\alpha_c \geq {\bar\rho}$. The lower bound is easily achieved, by a matrix inverse operation, when the positions of all zeros in the source are known -- or equivalently all the densities $\rho_i$ are extremal, $0$ or $1$. Any variation from these extremal values for fixed ${\bar \rho}$ increases the uncertainty in the source values, and so will increase the compression threshold.

The principal results in this article are regarding:
\begin{enumerate}
\item[(1)] The compression threshold $\alpha_c$ for perfect reconstruction, with high probability, given a set of weights;
\item[(2)] The assignment of weights, as a function of the marginal densities, that achieves a minimal $\alpha_c$, and the value of this minimum.
\end{enumerate}

\section{Analysis for w-$\ell_1$ minimization}
\subsection{Overview}
Rather than directly finding and classifying minima of the w-$\ell_1$ minimization problem, an auxiliary probability distribution may be introduced, for which the {\em maximum a posteriori} (MAP) solutions are equivalent to the solutions of the optimization problem. A parameter $\beta$ is introduced to soften the probability distribution and to allow application of analytic methods, and this is finally taken to be large so as to recover the posterior mean coincident with the MAP solution as well as the solution of the w-$\ell_1$ minimization.
The probability is the model posterior
\begin{equation}
P(\mathbf{\source}|\mathbf{y},\,A)= \frac{1}{Z}\delta\left(A\mathbf{\source}-\mathbf{y}\right)
\exp\left( -\beta \sum_i w_i |\source_i| \right) \label{eq:Px}
\end{equation}
where $Z$ is the normalization for the probability distribution.
In the limit of large $\beta$ the probability measure will be concentrated on values $\mathbf{x}$ that solve the weighted minimization problem (\ref{eq:wL1min}).
We further define a generating function in the large-system limit and its expectation value with respect to measurements
\begin{equation}
f = f(\beta,N) = -\frac{1}{\beta N}\log Z,\quad {\bar f} = \lim_{\beta\rightarrow\infty}\lim_{N\rightarrow\infty}\E_{A,\mathbf{y}}[f]\label{eq:f}\;.
\end{equation}
The function $f(\beta,N)$ is linearly related to a generalized version of the mutual information between
the source and the observations~\cite{Tanaka2002,GuoTanaka2010}.

Statistics on w-$\ell_1$ solutions are extracted from $f$, or ${\bar f}$, by adding small perturbations to the model posterior and taking derivatives.
Consider adding a small perturbation $-\beta h\|\mathbf{x}-\mathbf{x}^0\|_2^2$ to the exponent of (\ref{eq:Px}), whereby the normalization $Z$ now depends on $h$ as well.
Here, $\|\cdot\|_2$ denotes the $\ell_2$ norm.
The normalized mean square error (MSE) with respect to the model posterior (\ref{eq:Px}) is given by
\begin{equation}
N^{-1}\E[\|\mathbf{x}-\mathbf{x}^0\|_2^2] = \lim_{h\rightarrow 0} \frac{\partial}{\partial h} f(\beta,N)\;,
\end{equation}
which in the limit $\beta\to\infty$ gives the normalized MSE of the w-$\ell_1$ solution.
If MSE is evaluated to be zero, this implies that the w-$\ell_1$ minimization is sufficient for perfect reconstruction.

An analysis of $f$ is problematic owing to its definition in terms of random measurement parameters, but we mitigate for this by studying the asymptotics. Many interesting statistics, such as MSE, are found to be {\em self-averaging} -- the statistics derived from $f$ for almost all $\mathbf{y}$ and $A$ converge to those derived from ${\bar f}$ for large $N$.
We thus analyze the expectation value $\bar{f}$ in place of the random object $f$, expressing the former in an analytic form at leading orders in $N$ and $\beta$, thereby demonstrating conditions for perfect reconstruction.

In our analysis we assume, besides the self-averaging property, the validity of exchanging order of limits. We will solve the problem by the replica method, which introduces further heuristic assumptions - the {\em replica symmetric} (RS) solution is developed.  We are unable to verify rigorously each assumption, but the methodology is a standard one~\cite{Kabashima:TR,Rangan:AA,Nishimori:SP}, and a catalogue of successes indicates the results herein should be treated as exact. Some closely related results in compressed sensing~\cite{Kabashima:TR} are already known to coincide with those of rigorous methodologies~\cite{Donoho:CF}.

\subsection{The replica method}

The expectation with respect to the random variables (\ref{eq:f}) is taken by a standard approach. The generating function may finally be expressed in an extremization form, as
\WITHPRIOR{\begin{align}
{\bar f} &= \mathrm{Extr}_{\mraymond,\Qraymond,\chiraymond,\mconj,\Qconj,\chiconj} \left\lbrace \alpha \frac{\meanrho - 2 \mraymond + \chiraymond}{\Qraymond} + \mraymond\mconj - \frac{\chiraymond \chiconj}{2} + \frac{\Qraymond \Qconj}{2} \right.\nonumber\\
&+ \E_{\{\rho_i,w_i\}}\Bigr[\int \left((1-\rho_i)\delta(x^0) w_i^2 \epsilon(\lambda\sqrt{\Qconj}/w_i,\chiconj) \right.\nonumber\\
&+ \left.\left. \rho_i S(x^0) w_i^2  \epsilon((\sqrt{\Qconj} \lambda + \mconj x^0)/w_i,\chiconj)\right)\,D\lambda d x^0 \Bigr] \right\rbrace \label{eq:var_f}\;,
\end{align}}
\WITHOUTPRIOR{\begin{align}
{\bar f} &= \beta\mathrm{Extr}_{\mraymond,\Qraymond,\chiraymond,\mconj,\Qconj,\chiconj} \left\lbrace \alpha \frac{\meanrho - 2 \mraymond + \chiraymond}{\Qraymond} + \mraymond\mconj - \frac{\chiraymond \chiconj}{2} + \frac{\Qraymond \Qconj}{2} \right.\nonumber\\
&+ \E_{\{\rho_i,w_i\}}\Bigr[\int \left((1-\rho_i) w_i^2 \epsilon(\lambda\sqrt{\Qconj}/w_i,\chiconj) \right.\nonumber\\
&+ \left.\left. \rho_i w_i^2 \epsilon(\lambda\sqrt{\Qconj+\mconj^2}/w_i,\chiconj)\right)\,D\lambda\Bigr] \right\rbrace \label{eq:var_f}\;,
\end{align}}
where $D\lambda=(2\pi)^{-1/2}e^{-\lambda^2/2}d\lambda$ is a standard Gaussian measure, where $\E_{\{\rho_i,\,w_i\}}[\cdots]=N^{-1}\sum_{i=1}^N(\cdots)$ is an empirical average with respect to the densities and weights of source components, and where the function $\epsilon(\cdot,\,\cdot)$ is
defined as
\begin{equation}
\epsilon(a,b) = - \frac{(|a|-1)^2}{2 b}\Theta(|a|>1)\;.
\end{equation}
A solution is the extremum where the partial derivatives with respect to the six order parameters are zero, and the set of self-consistent equations describing the order parameters are called the saddle-point equations.
Note that for the special case of uniform weights\WITHPRIOR{ and Gaussian $S(x_i^0)$,} (\ref{eq:var_f}) becomes identical to results derived in the absence of marginal prior knowledge~\cite{Kabashima:TR}.

The order parameters for the correct solution of (\ref{eq:var_f}) coincide with informative quantities: $\chiraymond$ is the mean-squared estimate $N^{-1}\E[\|\mathbf{x}\|_2^2]$; $\mraymond$ is the overlap of source and estimate $N^{-1}\E[\mathbf{x}\cdot\mathbf{x}^0]$; and $\Qraymond$ is a statistic on the pair correlation functions. The mean square error is $N^{-1}\E[\|\mathbf{x}-\mathbf{x}^0\|_2^2]={\bar \rho} - 2\mraymond + \chiraymond$.

The six saddle-point equations are easily reduced to three equations on the conjugate parameters (denoted by hat) $\mconj$, $\chiconj$ and $\Qconj$, and further to two equations by identifying $\chiconj=\mconj$. Given a specific parameterization of the weights and measurement process these might be solved numerically. However, given our noiseless model (\ref{eq:yAx}) it is known that the mean square error will be zero in some range of large $\alpha$, and reasonable weights. This perfect reconstruction solution was found in the unweighted case with $1/\mconj=0$~\cite{Kabashima:TR}, and generalizes to the weighted case.

We can solve the equations to leading order in $1/\mconj$. At leading order many features of $S(x_i^0)$, the prior on non-zero components, are inconsequential.
It is convenient to define two functions of $\Qconj$
\begin{equation}
g_1(\Qconj) = \meanrho + \E[(1-\rho_i)2\mathcal{Q}(\wioversqrtQi)] \label{eq:f1}\;,
\end{equation}
where $\wioversqrtQi=w_i/\sqrt{\Qconj}$ is a rescaled weight, and
\begin{align}
g_2(\Qconj) &= \meanrho + \E[\rho_i (\wioversqrtQi)^2]
- \E\left[(1-\rho_i) \wioversqrtQi \frac{2\exp\left( - \wioversqrtQi^2/2 \right)}{\sqrt{2\pi}}\right] \nonumber\\
&+ \E\left[(1-\rho_i) (1 + (\wioversqrtQi)^2) 2\mathcal{Q}(\wioversqrtQi)\right] \label{eq:f2}\;,
\end{align}
where $\mathcal{Q}(z) = \int_{z}^\infty D\lambda$ is the conventional Q-function. The two saddle-point equations may be expressed at leading order as
\begin{equation}
\mconj^{-1} = \mconj^{-1}\left(\frac{\alpha}{g_1(\Qconj)}\right)^{-1} \label{eq:kappa}\;,
\end{equation}
and
\begin{equation}
\Qconj = \Qconj\frac{\alpha g_2(\Qconj)}{(g_1(\Qconj))^2} \label{eq:gamma}\;,
\end{equation}
where the latter equation can be solved numerically.

\subsection{Stability criteria}
For the RS solution of the replica method a local stability analysis gives necessary criteria for the validity of any solutions found, and can be studied by three eigenvalues derived from the full, rather than RS (\ref{eq:var_f}), saddle-point form~\cite{Almeida:SSK}. Two eigenvalues relate to instabilities consistent with the RS saddle-point equation (\ref{eq:var_f}), and can be evaluated within the RS framework, by analysis of (\ref{eq:kappa}) and (\ref{eq:gamma}).
The final 'replicon' eigenvalue can indicate a failure of the RS assumption yielding (\ref{eq:var_f}), but cannot be derived from (\ref{eq:var_f}). When the RS assumption fails in this sense, a symmetry breaking approximation should be required.

Stability within the RS framework requires that a small fluctuation in the order parameters of the right-hand side of (\ref{eq:kappa}) and (\ref{eq:gamma}) decays to zero under iteration of the equations. Small fluctuations in $\Qconj$ and $\mconj^{-1}$ about the perfect reconstruction solution are stable provided that
\begin{equation}
\alpha > g_1(\Qconj)\label{eq:stability}\;.
\end{equation}
It can be seen that for large $\alpha$ and small $\meanrho$ these conditions will be met. The replicon eigenvalue is consistent with the stability of the solution.

\section{Main results}
\subsection{Threshold equations}
Stability and uniqueness of the perfect reconstruction solution implies success of w-$\ell_1$ reconstruction up to errors of order $1/N$, with high probability. The instability criterion is met as $\alpha$ is decreased from one so that there exists a compression threshold $\alpha_c \leq 1$. A pair of equations describe the threshold in terms of only one order parameter $\Qconj$
\begin{equation}
\alpha_c =g_1(\Qconj)\qquad \hbox{and}\qquad g_3(\Qconj) = g_2(\Qconj)-g_1(\Qconj)=0 \label{eq:f3}\;.
\end{equation}
The latter equality is a consequence of (\ref{eq:gamma})
combined with (\ref{eq:stability}), and is independent of $\alpha_c$. Aside from the explicit dependence on $\Qconj$, there is a dependence on the densities and weights.
The two equations~(\ref{eq:f3}) provide implicit expressions
for the dependences of the compression threshold $\alpha_c$ on the set $\{\rho_i,\,w_i\}$ of densities and weights.

If one assumes uniform weights, the compression threshold $\alpha_c$ is described as an implicit function of $\bar{\rho}$ by
\begin{equation}
\alpha_c^{-1} = 1 + \sqrt{\frac{\pi}{2\Qconj}}\exp\left(\frac{1}{2\Qconj} \right) \left[1 - 2\mathcal{Q} \left(\frac{1}{\sqrt{\Qconj}}\right)\right]\label{eq:alphac}\;,
\end{equation}
and
\begin{equation}
\frac{\meanrho}{1-\meanrho} = \sqrt{\frac{2\Qconj}{\pi}}\exp\left(-\frac{1}{2\Qconj}\right)
-2\mathcal{Q}\left(\frac{1}{\sqrt{\Qconj}}\right)\;. \label{eq:meanrho}
\end{equation}
For an inhomogeneous system of fixed $\meanrho$, any knowledge of the marginal densities constitutes additional information. Incorporation of non-uniform weights introduces additional degrees of freedom in the model that can be exploited to reduce $\alpha_c$. It also means that it is necessary to specify the weights according to the given set of densities in order to find the optimum value of $\alpha_c$.

\subsection{Optimization of $\alpha_c$ by weight selection}
In order to select optimally the weights, we consider the optimization problem of minimizing the compression threshold with respect to the weights. For this purpose we can take $\Qconj$ and $\{\rho_i\}$ to be fixed parameters and minimize $\alpha_c=g_1$ with respect to the weights $\{w_i\}$ subject to the constraint that $g_3$ is zero (see (\ref{eq:f3})). The constraint can be dealt with via the Lagrange multiplier method as
\begin{equation}
\min_{\{w_i\}, \lambda}\left( g_1 - \lambda g_3\right)\;.
\end{equation}
The derivative with respect to $w_i$ leads to the following
criteria
\begin{equation}
\wioversqrtQi \exp\left( \wioversqrtQi^2/2\right) \left(\frac{\rho_i}{1-\rho_i}+2\mathcal{Q}(\wioversqrtQi)\right) = \frac{1+\lambda}{\lambda\sqrt{2\pi}} \label{eq:wb}\;,
\end{equation}
combined with the derivative with respect to $\lambda$, which reads
\begin{equation}
\frac{1}{\sqrt{2\pi}} \E\left[\wioversqrtQi(1-\rho_i) \exp(-\wioversqrtQi^2/2)\right] \left(\frac{1+\lambda}{\lambda}- 2\right) = 0\label{eq:lambda}\;,
\end{equation}
yielding $\lambda=1$. Therefore the set of equations (\ref{eq:wb}), each of which determines one weight, are independent given $\Qconj$ and $\rho_i$.

Using this independence it is possible to optimally set the weights for an arbitrary density distribution $\{\rho_i\}$ by the following procedure. Solve (\ref{eq:wb}) for $\wioversqrtQi$, and the solution is a function of $\rho_i$ only. The set of solutions $\{u_i\}$ defines the optimal weights up to the overall scaling. Observing that the w-$\ell_1$ minimization problems are invariant under the overall scaling, every weight can be assigned straightforwardly as $w_i=\wioversqrtQi$. This set of weights will minimize $\alpha_c$.

To evaluate the minimum of $\alpha_c$, one observes from (\ref{eq:f1}) and (\ref{eq:f3}) that $\alpha_c=\E[\alpha(\rho_i)]$ holds, where $\alpha(\rho_i)=\rho_i+(1-\rho_i)2\mathcal{Q}\bigl(u_i(\rho_i)\bigr)$.
This expression can be understood as if each component of the source, with density $\rho$, would require a compression rate at least $\alpha(\rho)$ for its perfect reconstruction, and the lower bound of the total compression rate is the empirical average of this componentwise bound over all components. This interpretation would further suggest a possibly fundamental importance of the quantity $\alpha(\rho)$ as a measure of some sort of information associated with a random variable that takes non-zero values with probability $\rho$. One can also confirm, by particularizing our results to the unweighted system, that $\alpha(\rho)$ gives the compression threshold of the unweighted system with density $\rho$. Thus, evaluation of $\alpha_c$ for the weighted system only requires the densities $\{\rho_i\}$ of the system and the curve $\alpha(\rho)$ for the unweighted system. Moreover, the curve for the unweighted system is convex upward, so that Jensen's inequality tells us that $\alpha_c$ is below $\alpha(\meanrho)$, except when the density distribution is concentrated at a single point, implying that the w-$\ell_1$ minimization improves the compression threshold over the unweighted counterpart.

It should also be noted that, when a block structure can be assumed for the source, the minimum $\alpha_c$ is equal to the compression rate that would be achievable if each block were measured separately with the optimum compression rate for that block and then reconstructed individually from other blocks. This observation implies that introduction of optimal weights successfully compensates degradation due to intermixing of blocks of different densities in measurements.

Finally we can note an interesting asymptotic in the result applicable in the case that $\meanrho$ is close to one ($\Qconj$ small). In this case the optimal weights are assigned according to a simplified form of (\ref{eq:wb})
\begin{equation}
\rho_i/(1-\rho_i) \sim \sqrt{\frac{2}{\pi}}\frac{\exp\left(
-\wioversqrtQi^2/2\right)}{\wioversqrtQi^3}.
\end{equation}

\section{A two-block example}
\subsection{Setting}
As a simple case we consider the source consisting of two equally-sized blocks labeled $b=\pm 1$, such that components in the block $-1$ are non-zero with some uniform lower probability than components of block $+1$. A corresponding asymmetry in the weights can be assumed, but weights must also be uniform within any block. This case allows a deal of intuition and has been studied previously owing to its simple structure~\cite{Xu:BT}.

\subsection{The large-system limit}
\begin{figure}[!t]
\centering
\includegraphics[width=\linewidth]{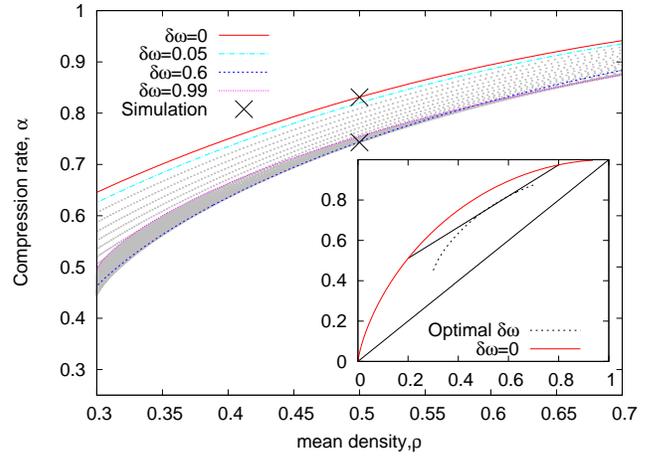}
\caption{(color online) Compression threshold $\alpha_c$ as a function of density $\{0.3<\meanrho<0.7,\ \delta \rho = 0.3\}$ for various $\delta w$. Four curves are highlighted from a range $0\leq\delta w\leq0.99$. The lower hull of all curves indicates the minimum $\alpha_c$ achievable by optimal weight selection. Inset: The hull of all curves (corresponding to the maximum compression) can be constructed from the unweighted ($\delta w = 0$) curve by a simple linear construction.}
\label{fig:variousweights}
\end{figure}

Let the density and the weight of block $b=\pm1$ be $\rho_b=\meanrho+ b\,\delta\rho$ and $w_b=1 - b\,\delta w$, respectively, with $\delta w,\,\delta\rho\ge0$. Figure~\ref{fig:variousweights} demonstrates the result for a variety of weight asymmetry $\delta w$ given a density asymmetry $\delta\rho=0.3$. Since the density in each block must be in $(0,1)$ the relevant values of $\meanrho$ are confined to the interval $(0.3,0.7)$.
Curves for $0\leq\delta w\leq0.99$ are plotted, with four labeled cases, the uppermost curve being the unweighted case. It can be seen that the compression threshold is reduced by allowing asymmetric weights, and that the lower envelope of all curves indicates the achievable performance by optimal selection amongst the weights.
For the system $(\rho_{+1},\,\rho_{-1})=(0.8,\,0.2)$
the midpoint of the line connecting $(\rho_{+1},\,\alpha(\rho_{+1}))$ and $(\rho_{-1},\,\alpha(\rho_{-1}))$
coincides with the lower envelope at $\meanrho=0.5$ corresponding to the optimal compression rate
for the two-block model.

\subsection{Numerical verification}
\begin{figure}[!t]
\centering
\includegraphics[width=\linewidth]{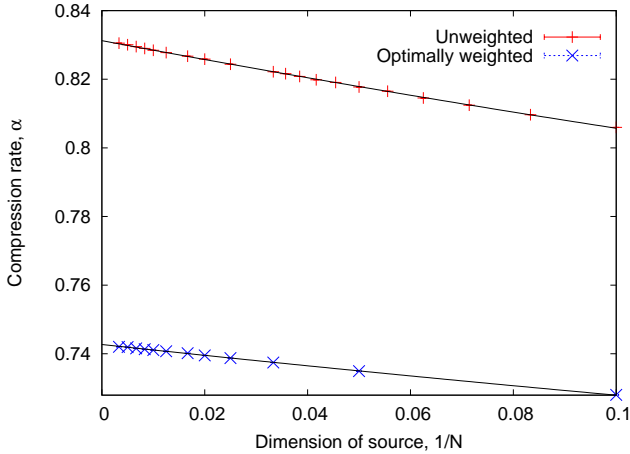}
\caption{(color online) Compression threshold $\alpha_c$ for unweighted ($\delta w=0$) versus optimally weighted ($\delta w =0.684$) reconstruction for a two-block model ($\rho_{+1}=0.8$, $\rho_{-1}=0.2$) for various system sizes. Each symbol represents the mean threshold from $10^{6}$ trials each with independently generated measurements, error bars are small by comparison with symbol size. The data is fitted by $\chi^2$ regression.}
\label{fig:optimalweights}
\end{figure}

To test the theory we have generated a number of instances of the problem for various system sizes
and solved these by linear programming.
Components of $A$ and non-zero components of $\mathbf{x}^0$ were sampled from normal distributions.
Optimally weighted, and unweighted, $\ell_1$ minimization problems were solved
to determine the compression threshold where optimal reconstruction failed.
Figure~\ref{fig:optimalweights} demonstrates result for a two-block system
with a uniform weighting $w_{+1}=w_{-1}$ and optimized weighting.
It can be seen that the thresholds are distinct with a significant improvement
in performance for the optimally weighted case.

Results of fitting these numerical data with second-order polynomials in $1/N$, obtained by $\chi^2$ regression, are also plotted in Fig.~\ref{fig:variousweights}. Extrapolation to $1/N\to0$ yields an estimate of the compression threshold $\alpha_c=0.742\,73\pm0.000\,06$ for the optimally weighted case, which is in agreement with the analytical result $\alpha_c=0.742\,72$. The extrapolated result for the unweighted case $\alpha_c=0.831\,32\pm0.000\,06$ is also in agreement with the analytical result $\alpha_c=0.831\,30$.

\section{Conclusion}
This paper has demonstrated a method for optimal selection of weights in the w-$\ell_1$ minimization utilizing prior knowledge about densities, and thereby providing the optimal threshold for compression. The result is a simple one with decoupling structure for source components, and described by a single order parameter in the case of perfect reconstruction. The threshold in the compression rate for which perfect reconstruction is possible in a system of known marginal densities can be straightforwardly derived from the threshold curve for the unweighted system with a simple graphical procedure. This work should in future be extended to consider the effect of noise and of correlations between the source components.

The method relied on the replica method and a saddle-point formulation, which although complicated in origin provides a concise and intuitive saddle-point framework from which to derive results.

The analysis presented has been verified in experiment and provides a mechanism that may be immediately incorporated in practical problems where marginal density information is available. A complete description of the replica analysis and various extensions will be forthcoming in an article under preparation.

\section*{Acknowledgment}
Support from the Grant-in-Aid for Scientific Research on Priority Areas ``Deepening and Expansion of Statistical-Mechanical Informatics'' by the Ministry of Education, Culture, Sports, Science and Technology, Japan (No.~18096010), is acknowledged. J.R. is supported in part by Research Grants Council of Hong Kong (Grant No.~HKUST 604008,~603607).

\BIBFILEYES{
\bibliographystyle{IEEEtran}
\bibliography{IEEEabrv,Bibliography_20100408}
}
\BIBFILENO{

}

\end{document}